\documentstyle[psfig]{elsart}

\begin{document}
\runauthor{Rosa M. Gonz\'alez Delgado}
\begin{frontmatter}
\title{The stellar population and the evolutionary state of 
HII regions and starburst galaxies}

\author{Rosa M. Gonz\'alez Delgado}

\address{Instituto de Astrof\'\i sica de Andaluc\'\i a (CSIC), 
Apdo. 3004, 18080 Granada, Spain. e-mail:rosa@iaa.es}

\begin{abstract}

RH\,{\sc ii}s and starbursts are both powered by massive stars. 
They are the main contributors to the heating of the ISM via 
radiative and mechanical energy. Techniques to derive the
stellar content and the evolutionary state of 
RH\,{\sc ii}s and starbursts from their ultraviolet and optical integrated light
are reviewed. A prototypical RH\,{\sc ii} (NGC 604) and nuclear 
starburst (NGC 7714) are discussed in more detail. The results reveal the 
necessity of multiwavelength analyses of these objects to estimate their stellar content and 
their evolutionary state in a consistent way.


\end{abstract}

\end{frontmatter}

\section{RH\,{\sc ii}s and starbursts: Definition and common characteristics.}

Giant extragalactic H\,{\sc ii} regions (RH\,{\sc ii}s) are amongst the
brightest objects in galaxies. They have been studied extensively
because they are the best indicators of the conditions that lead to massive
star formation, and they show the cloud properties inmediately after the moment
when stars form. RH\,{\sc ii}s are characterized to have a size larger than 100 
pc and H$\alpha$ luminosity brighter than 10$^{39}$ erg s$^{-1}$ \cite{ken}.
Thus, the nebula requires an ionizing photon luminosity $\geq$ 10$^{51}$ ph~s$^{-1}$; 
this is provided by a stellar cluster that contains more 
than 100 young massive stars. These characteristics are very similar to those of
starburst galaxies. However, they are less luminous than prototypical
starbursts, and thus, they can be referred to as mini-starbursts \cite{wal91}. 

Starburst galaxies are objects in which the total energetics are dominated by 
star formation and associated phenomena \cite{weedman}. They are 
characterized to have a size between 100 pc and 1000 pc, and H$\alpha$ luminosity 
that ranges between 10$^{40}$ erg s$^{-1}$ and 10$^{42}$ erg s$^{-1}$.
Therefore, the nebula requires an ionizing photon luminosity that ranges between 
10$^{52}$ ph~s$^{-1}$ and 10$^{54}$ ph~s$^{-1}$; this is provided by a stellar 
cluster that contains several thousands of young massive stars.
This definition covers galaxies with a very wide range of properties such as: blue compact 
dwarfs, H\,{\sc ii} galaxies, nuclear starbursts and ultraluminous IRAS starbursts. 
Typical bolometric luminosities range from 10$^7$ to 10$^{12}$ L$\odot$, corresponding 
the lowest limit to the luminosity of super-star-clusters and the highest limit 
to infrared-luminous galaxies \cite{lei96}. The star formation rate is 
so high (10--100 M$\odot$ yr$^{-1}$ and in some ultraluminous infrared galaxies can be up to
1000 M$\odot$ yr$^{-1}$) that the existing gas supply can feed the starburst only
for a small fraction of the age of the universe (few 10$^8$ yr). 

One common definition to RH\,{\sc ii}s and starbursts is: {\em brief intense star formation 
episodes that are taking place in small regions}. The main 
difference between both is that the luminosity of these episodes of star formation 
dominate the overall luminosity of the starburst galaxy (L$_{burst}\sim$ L$_{galaxy}$),
in contrast in RH\,{\sc ii}s is significant smaller (L$_{burst}\ll$ L$_{galaxy}$) 
\cite{terl}.

A common characteristic to both is their ultraviolet and optical spectral morphology.
RH\,{\sc ii}s and starbursts are powered by massive stars. O stars emit photons with 
energies of tens of eVs that are absorbed and re-emitted in their stellar winds, 
producing ultraviolet resonance transitions. These wind lines are blueshifted by 
2000--3000 km s$^{-1}$ and/or show a P-Cygni profile \cite{groenewegen}. The shape of the 
profiles
reflect the stellar mass-loss rates which are a strong function of the stellar luminosity 
\cite{castor}.
Therefore, the ultraviolet spectrum of RH\,{\sc ii}s and starbursts is dominated by
absorption lines formed in the wind of massive stars 
\cite{rosa,kinney}. However, it can also show weak 
absorption lines formed in the photosphere of O and B stars \cite{mello}, and 
strong absorption lines formed in the interstellar medium of the
galaxy \cite{heck97,gd98}. In contrast, the optical spectrum is dominated by 
nebular emission lines. 
The stellar wind is optically thin to most of the ultraviolet photons, 
that can travel tens of parsecs from the star before they are absorbed and 
photoionize the surrounding interstellar medium. Subsequently, this ionized
gas cools down via an emission line spectrum. However, at optical 
wavelengths, the starbursts and some RH\,{\sc ii}s show stellar features formed in the 
photosphere of early type stars (e.g. the higher order Balmer series and HeI lines) 
\cite{terl96,storchi}. 
Other stellar lines detected
in starburst are at near-infrared. The stronger one are the CaII triplet at 8600 \AA\ and 
the CO bands at 2.2 $\mu$ \cite{oliva}. These lines forme in the photosphere of giant 
and supergiant
stars. However, these lines, that are detected in most of the starbursts, are
very difficult to detect in RH\,{\sc ii}s, probably because they have very little 
red-supergiant stars (see e.g. \cite{terl96} and \cite{drissen}). 

The spectral morphology of RH\,{\sc ii}s and starbursts at the ultraviolet and 
optical wavelengths allow to derive the stellar content and 
the evolutionary state of the cluster in a self-consistent way by doing multiwavelength
analysis of these objects. 
In this paper I will give an overview of different techniques to date RH\,{\sc ii}s and 
starbursts and to estimate their stellar content. For guiding the discussion
I will apply these techniques to the prototypical nuclear starburst NGC 7714 and the 
prototypical RH\,{\sc ii} NGC 604. The ultraviolet 
and optical apperance of these objects are shown in Figures 1 and 2. However, first 
I will discuss the effect that extiction has on the ultraviolet emission. 

\setcounter{figure}{0}

\begin{figure}
\psfig{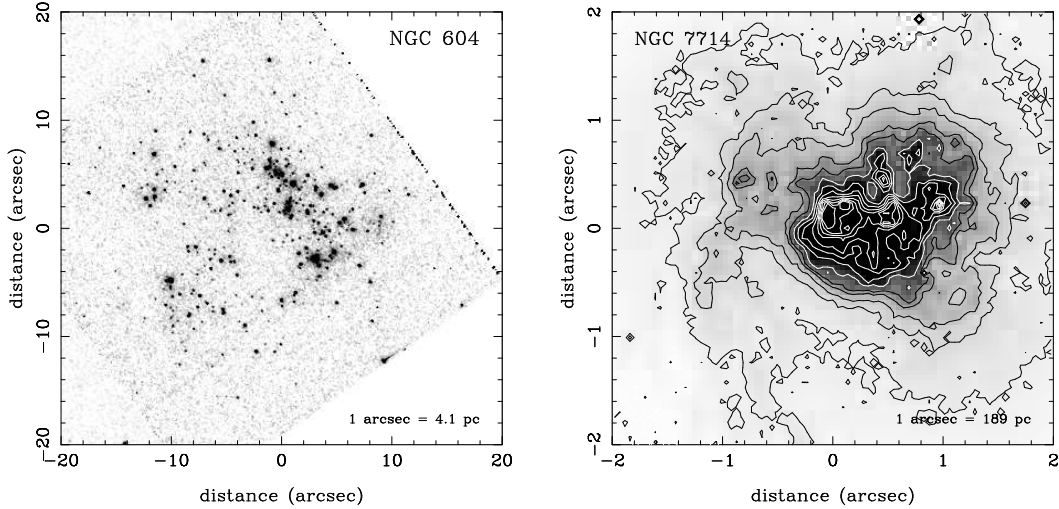}
\caption{{\it HST\/}+WFPC2 image of central stellar cluster of NGC 604 at 1700 \AA\ (left) 
and the nucleus of NGC 7714 at U band (right).} 
\label{HSTuv} 
\end{figure}

\begin{figure*}
\psfig{figure=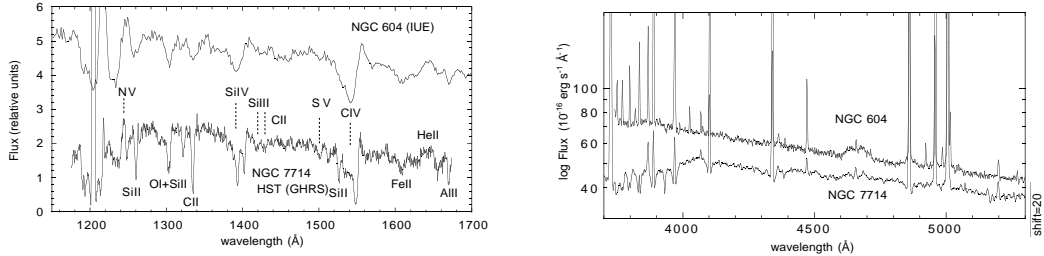,width=150mm,angle=0}
\caption{Ultraviolet (left) and optical (right) spectra of NGC 604 
and of the nucleus of NGC 7714. The spectra show the dichotomy picture 
characteristic of starbursts and RH\,{\sc ii}s: The UV wavelengths 
dominated by absorption lines and the optical wavelengths by nebular emission lines.
Broad HeII $\lambda$4686 emission is detected in the optical spectra of both objects.} 
\label{uvopt} 
\end{figure*}

\section{Extinction}

One way to derive the massive stellar content in RH\,{\sc ii}s and starbursts is through the ultraviolet
continuum emitted by these objects. However, due to the profound effect that dust has on the 
ultraviolet properties of starbursts, an estimation of the extinction needs to be done first.
\cite{schmitt} have built the spectral energy distribution from X-ray to radio of 
a sample of ultraviolet local starbursts. They divided the sample in two classes, low-reddening
(E(B-V)$\leq$ 0.4) and high-reddening (E(B-V)$\geq$ 0.4). Both have similar spectral energy
distribution ({\sc sed}) over the entire energy spectrum, peaking at the far-infrared. However, 
low-reddening starbursts have stronger ultraviolet emission than high-reddening starbursts, 
while in the far-infrared the opposite happens. This difference is due to the fact that 
the ultraviolet and visual radiation that is absorbed by dust is re-radiated in the far-infrared.
This suggests that dust reddening is the likely cause for extinguished the ultraviolet spectra. Thus, 
the extinction has to be estimated in order to obtain the ultraviolet intrisic luminosity of the starburst.
\cite{calzetti} show that in a sample of local starburst galaxies, the ultraviolet 
continuum can be paremetrized as F$_\lambda$= cte $\lambda^\beta$, with the spectral 
index $\beta$ strongly correlating with the nebular extinction measured using the Balmer 
decrement. This spectral index also correlates with the ratio of the far-infrared to 
the ultraviolet flux, $L_{IR}/L_{UV}$, which is larger for $\beta$ values \cite{meurer}. 
\cite{heck98} have
also shown that $\beta$ and $L_{IR}/L_{UV}$ ratio correlate with the metallicity of the gas,
the strength of the ultraviolet interstellar and wind absorption lines. These correlations 
point out that the more metal-rich starbursts are redder and more heavily extinguished in 
the ultraviolet.

Fortunately, $\beta$ is independent of the IMF and the 
star formation history, and it ranges 
between -2.0 and -2.6, for ages appropriate for starbursts \cite{lei95a}.
This allows to do an estimation of the extinction that affect the stellar cluster
by comparing the observed ultraviolet continuum with the spectral
energy distribution predicted by the evolutionary synthesis models. Then, the ultraviolet continuum is 
de-reddened to get the intrinsic ultraviolet luminosity and the massive stellar content. The color excess, E(B-V),
derived in this way for local ultraviolet selected starburst galaxies is usually lower
 than 0.4 \cite{mas}. In particular, the values estimated for NGC 604 and NGC 7714 are 0.1 and 
0.03, respectively.
However, E(B-V) derived in this way is in many starbursts significantly smaller (by a 
factor 2) than that derived from the Balmer decrement \cite{fanelli,calzetti}. This result, 
therefore, suggests that the ionized gas is associated to regions of higher dust content 
than the stars. The most probably 
explanation is that the dust has been swept away and/or destroyed from the site of
star formation by the action of the stellar winds and supernova explosions. 
One important implication of 
this result is that the equivalent widths of the Balmer recombination lines 
are not free extinction quantities; thus, this limit their use as age diagnostic.

\section{Evolutionary state of RH\,{\sc ii}s and starbursts}

One important question related with RH\,{\sc ii}s and starbursts is whether the star formation 
proceeds in them in short (duration $\leq$ 10 Myr) or long lived bursts (duration $\geq$ 10 Myr).  
There are several different diagnostics that can be used to determine the 
evolutionary state of RH\,{\sc ii}s and starbursts which are all based on their spectral morphology.
Here, I give an overview of several techniques
to date star forming systems which are based on the strength of the nebular
emission lines, wind resonance ultraviolet lines, H Balmer series and HeI absorption 
lines, and the spectral energy distribution ({\sc sed}) from the ultraviolet to the near infrared.
I present the results of applying these techniques to NGC 604 and NGC 7714 (see 
\cite{gd20} and \cite{gd99a} for more details).

\subsection{Nebular emission lines}

The emission line spectrum of an RH\,{\sc ii} and starburst depends on the radiation field 
from the ionizing stellar cluster, on the electron density and on the chemical 
composition of the gas. A photoionization code can take as input the spectral 
energy distribution of the cluster, and solve the ionization-recombination 
and heating-cooling balances to predict the ionization structure of the nebula, 
the electron density and the intensity of the emission lines. 
The star formation law, age and massive stellar content of the stellar 
clusters can be constrained by comparing the observed emission lines strengths 
with the predictions from the photoionization models, if the code uses as 
input the {\sc sed} generated by a stellar evolutionary 
synthesis code. This technique has been used successfully to study the stellar 
content in starbursts and RH\,{\sc ii}s (e.g. \cite{garcia,stas96} 
and Stasi\'nska this conference). However, the age range for 
which this technique is sensitive is restricted to the first 10 Myr, that is the evolutionary 
timescale of O stars. The thruthfulness of the results obtained with this technique depends 
strongly on the observational constraints used for the modeling. Important limitations to 
this technique comes from 
the effect of dust on the ionizing radiation field and of the geometric distribution of 
the gas on the ionizing structure of the nebula \cite{stas99}. Thus, constraints on the 
spatial distribution of the ionized gas obtained from high resolution narrow-band images 
are very useful. The success of this technique relies on obtaining ages and information 
on the {\sc sed} 
of the stellar cluster that are compatible with those resulting from other techniques
like the modeling of the wind resonance ultraviolet lines.

To determine the evolutionary state of the prototypical RH\,{\sc ii} NGC 604 and nuclear 
starburst NGC 7714, the nebular lines were predicted by the photoionization code 
{\sc cloudy} \cite{ferland}. 
In these models the radiation field is the spectral energy distribution from the 
evolutionary synthesis code Starburst 99 \cite{lei99} that is normalized to 
the ionizing photon luminosity derived from the 
total H$\alpha$ flux of the nebula. The electron density and chemical composition of 
the gas are fixed to the values derived from the observations. The gas is distributed 
with constant density in a sphere with an inner radius of a few parcsecs and an outer 
radius that is determined by the ionizing front. The models assume that the gas
occupies only a fraction of the sphere; then, they are computed taking as a free 
parameter the filling factor, $\phi$, that is related to the average ionization parameter U by
U= $(\phi^2~N_e~Q)^{1/3}$. First, U is derived using the ratio [S {\sc ii}]6717+6732/H$\beta$
that is a good calibrator of U \cite{gd99a}. The observed ratios indicate $\phi$ of
0.1 and 0.001 for NGC 604 and NGC 7714, respectively. These models predict a Str\"omgren radius
of 110 pc and 185 pc for NGC 604 and NGC 7714, respectively. These values are in agreement with 
the nebular radius estimated from H$\alpha$ images of these objects. 
Then, the strength of other emission lines 
are fitted. Line ratios indicative
of the electron temperature and ionization structure suggest that the massive stellar
population responsible for the photoionization of the gas has formed in an instantaneous 
burst 3 and 4.5 Myr ago in NGC 604 and NGC 7714, respectively (Figure 3 and 4).

\begin{figure}
\psfig{figure=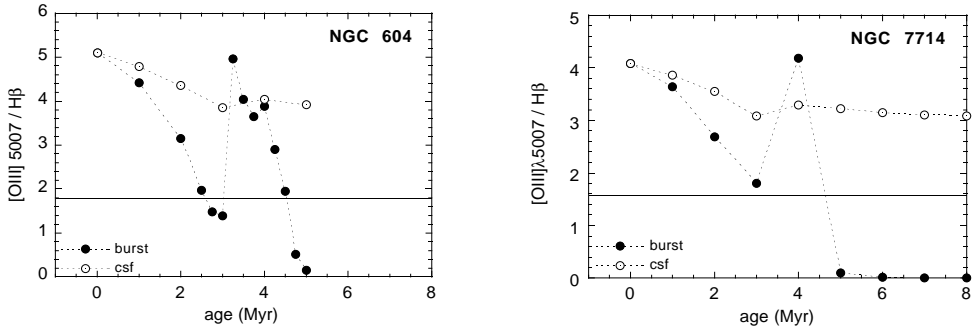,width=145mm,angle=0}
\caption{Predicted [O {\sc III}]/H$\beta$ by {\sc cloudy} using the {\sc sed}
from Starburst 99. The observed values are marked as horizontal lines. Models shown are those 
computed for $\phi$= 0.1 (for NGC 604) and 0.001 (for NGC 7714). The electron density is 30 and 100 
cm$^{-3}$ for NGC 604 and NGC 7714, respectively. The chemical composition is Z$\odot/2$.} 
\label{ratio} 
\end{figure}

\begin{figure}
\psfig{figure=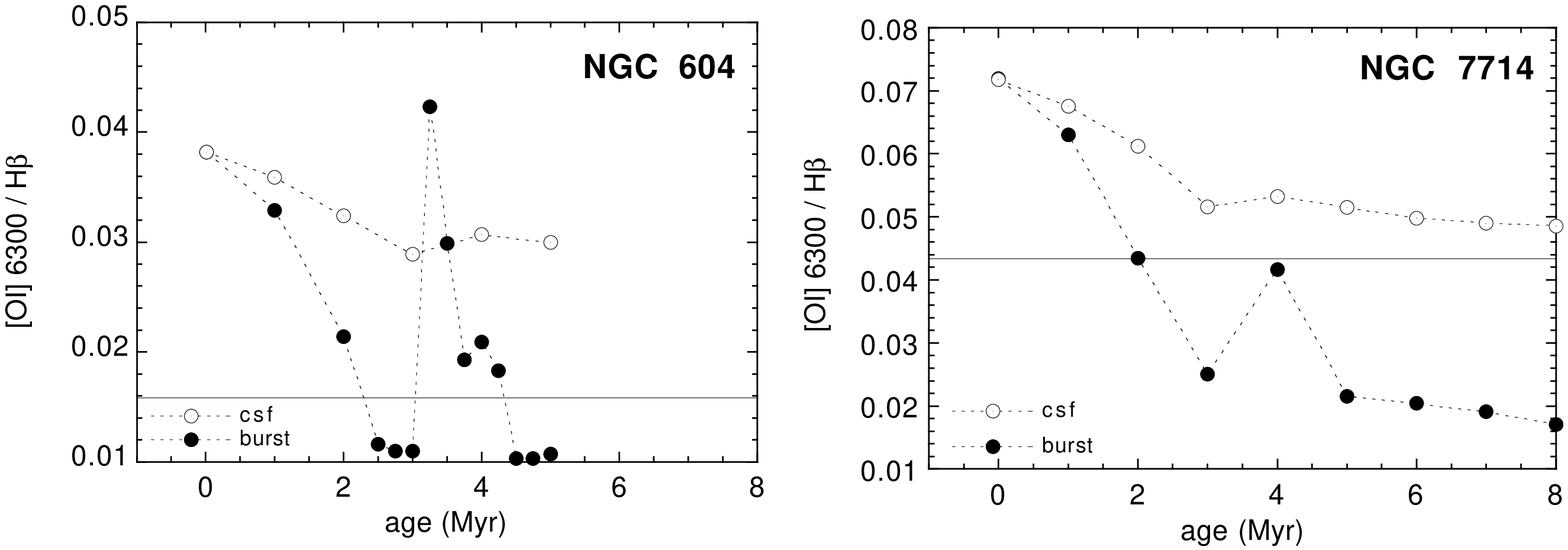,width=145mm,angle=0}
\caption{As Figure 3 for the line ratio [O {\sc I}]/H$\beta$.} 
\label{ratio} 
\end{figure}

\subsection{Ultraviolet stellar lines}

Stellar winds are driven by radiation pressure \cite{morton}. 
Massive stars transform their radiative 
momentum into kinetic energy with an efficiency $\eta$= (\.M v$_\infty$)/ (L/c) $\sim$ 0.3 
for O stars of solar metallicity \cite{lamers}, where \.M is the mass loss rate, v$_\infty$ 
the terminal velocity of the wind, L the luminosity of the star and c the light speed. 
This expression explains why the profile of the wind resonance ultraviolet lines
contains information about the mass-loss rate and on the stellar luminosity. Since there is 
a well-defined stellar mass-luminosity relation, the profiles intimately deppend on the
stellar content of the starburst and its evolutionary state. The most prominent wind lines
in the spectra of RH\,{\sc ii} and starburst are those of N\,{\sc v} $\lambda$1240, Si\,{\sc iv} $\lambda$1400, 
C\,{\sc iv} $\lambda$1550 and He\,{\sc ii} $\lambda$1640. C\,{\sc iv} shows always a strong 
P-Cygni profile if O stars with zero-age-main-sequence masses above 30 M$\odot$ are present.
In contract, Si\,{\sc iv} and He\,{\sc ii} forms if blue supergiant stars and stars with very 
dense wind (as
Wolf-Rayet) are present in the cluster (respectively). Evolutionary synthesis models show 
that the profile of these lines depend on the IMF parameters and the evolutionary state 
of the starburst \cite{lei95a}. As the nebular emission lines, 
the age range for which this technique is sensitive is only the first 10 Myr. After this age, 
the cluster is dominated by B stars and photospheric lines of C and Si can be
used to estimate the evolutionary state of the cluster \cite{mello}.
One important limitation of this technique comes from the strong dependency that the 
stellar winds have with the metallicity and the difficulty to built a stellar
library at sub-solar metallicity. 

The strength of the wind lines in NGC 604 and NGC 7714 indicate that the stellar cluster
formed in an instantaneous burst 3 Myr and 5 Myr ago, respectively (Figure 5). These ages are
consistent with the presence of Wolf-Rayet stars, as indicated also by the wind lines 
He\,{\sc ii} $\lambda$1640 and $\lambda$4686 (see Figure 2).
In both objects, stars more massive than 80 M$\odot$ formed in the cluster and they follow a 
Salpeter IMF (or even flatter IMF in the case of NGC 604). This result is consistent with
 that obtained from the analysis of the nebular emission lines. Therefore, we conclude that
the stellar cluster responsible of the photoionization of the interstellar gas and the 
ultraviolet continuum emission in these two prototypical objects formed in a short period of
time a few Myr ago.
 
\begin{figure*}
\psfig{figure=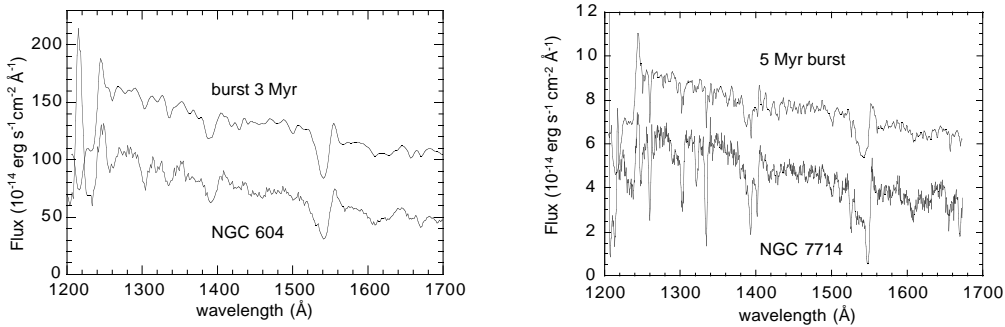,width=145mm,angle=0}
\caption{De-reddened ultraviolet spectra of NGC 604 (left) and the nucleus of NGC 7714 (right) 
and the synthetic burst models (shifted in flux for clarity) that fit the wind stellar lines.} 
\label{uvfit1} 
\end{figure*}

\subsection{H Balmer and HeI photospheric lines}

Evidence that other stellar populations contribute significantly to the optical and near-infrared
wavelengths come from other diagnostics that are sensitive to the presence of B and A stars, as for 
example the H Balmer series and He\,{\sc i} absorption lines.
The spectra of early type stars are characterized by strong H Balmer and He\,{\sc i} 
absorption lines, with very weak metallic lines formed in the photosphere of O, B and A 
stars \cite{wal90}. In the spectra of RH\,{\sc ii}s and starbursts 
the H Balmer and He\,{\sc i} recombination 
nebular emission lines are super-imposed on the corresponding photospheric lines.
However, the higher order terms of the Balmer series and some of the He\,{\sc i} 
lines can be detected in absorption or show absorption wings. These stellar lines 
can be detected easier in absorption than the lower terms of the Balmer series, as H$\alpha$, 
because the strength of the gaseous Balmer lines in emission decreases rapidly with 
decreasing wavelength, whereas the equivalent width of the stellar absorption lines is
constant with wavelength \cite{gd99b}. These lines are very sensitive
to the age if the burst is younger than a few Myrs and can be used to date starburst
and post-starburst galaxies until 1 Gyr \cite{gd99b}. 

In very young stellar systems (few Myr old),
these lines can also be detected in absorption
in spectra where the continuum light is maximized with respect to the nebular contribution.
For example, a spectrum corresponding to the central zone ($\sim$ 5$\times$5 arcsec) of NGC 604
can show the higher order terms of H Balmer series and He\,{\sc i} lines in absorption because 
the core of the stellar cluster is in a hole of nebular emission. The age 
of the stellar population that dominates the optical light can be estimated comparing 
the strength of these lines with the prediction of the evolutionary synthesis models. 
For NGC 604, the shape of these lines is compatible with the profiles predicted by an 
instantaneous burst 
3 Myr old (Figure 6). However, in NGC 7714 these lines are stronger and 
wider than those predicted by the instantaneous 
burst, 4-5 Myr, that fits the ultraviolet continuum and the nebular emission lines. 
An older population ($\sim$ 200 Myr old) contributes significantly to the optical light 
of the nuclear starburst of NGC 7714 (Figure 6).
  
\begin{figure*}
\psfig{figure=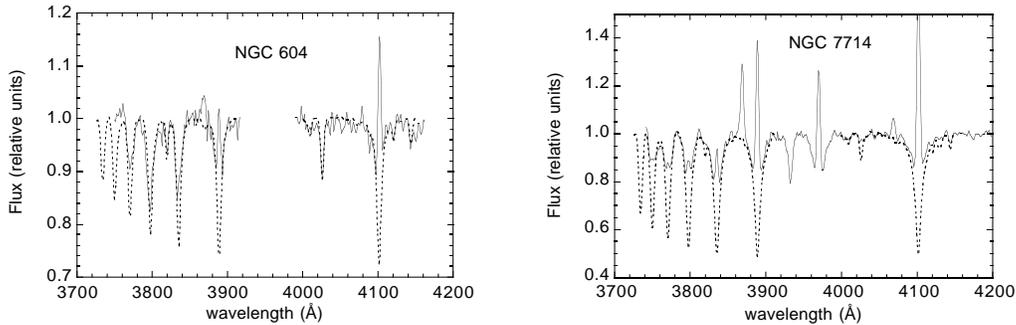,width=145mm,angle=0}
\caption{Normalized optical spectrum of inner 1.2$\times$4.6 arcsec of NGC 604 (left) fitted by 
a burst 3 Myr old (dotted line); and the nucleus of NGC 7714 (right) fitted by a composite model with contributions of 
burst of 5 Myr and 200 Myr old (dotted line).} 
\label{uvfit1} 
\end{figure*}
   
\subsection{SED from the ultraviolet to the near-infrared}

Evidence that starbursts experience several recurrent bursts of star formation also comes
from the analysis of the spectral energy distribution from the ultraviolet to the near-infrared. 
A composite ultraviolet-optical-near-infrared spectrum of the inner 2 arcsec ($\sim$ 380 pc)
of the nucleus of NGC 7714 indicate that an instantaneous burst 4.5 Myr old that fits the 
ultraviolet light contributes less than 30$\%$ to the optical and less than 5$\%$ to
the near-infrared light. As the strength of the Balmer absorption lines indicates an intermediate
population, $\sim$ 200 Myr, contributes significantly to the optical light. However, these 
two populations can not account for the strong near-infrared continuum detected in NGC 7714. An
additional population rich in red supergiant stars, $\sim$ 10 Myr, heavily obscured is needed
to fit the near-infrared light; the ultraviolet flux produced by this population is totally
hidden from view (Figure 7) \cite{goldader}. This result is in the line of that found by 
\cite{heck98} that a significant fraction of the stellar population of a starburst is obscured,
because even 
after correcting by extinction the ultraviolet flux represent only 10$\%$ of the 
actual luminosity of the starburst.
In contrast, the spectral energy distribution of NGC 604 is well fitted by an instantaneous burst 
3 Myr old (Figure 7).
Therefore, no evidence of recurrent star formation episodes is found in the inner 20$\times$20 
arcsec of NGC 604.

Thus, the star formation in the prototypical starburst galaxy NGC 7714 proceeds
in a very different way to that in the prototypical RH\,{\sc ii}s NGC 604. Thus while, in the 
starburst the star formation is taking place for long period of time through several recurrent bursts, 
in more simple systems, as NGC 604, the star formation lasts for a very short period of time. 
However, evidence
that multiple stage starburst occurs in other simple stellar systems exists. For example, in the 
RH\,{\sc ii} NGC 2363 two burst of less than 1 Myr and $\sim$ 3 Myr old is taking place 
over a linear scale of 100 pc \cite{drissen}. Other good example is in the 
super star cluster A (SSC A) of NGC 1569. Its
ground-based optical spectra show Wolf-Rayet features at 4686 \AA\ and near-infrared CaII
triplet at 8600 \AA\ in absorption. Its spectral energy distribution is well fitted by two burst model, with
the younger burst having an age of 3 Myr and the older having $\sim$ 10 Myr, to explain the
simultaneous presence of hot massive stars and red supergiants \cite{gd97}.
This result is supported by HST optical images that reveal that SSC A is resolved in two 
subclusters separated by a projected distance of $\sim$ 2 pc (\cite{marchi}).
Even these two subclusters would not be connected, a comparison with the star formation scenario in
30 Doradous (the central cluster R136 is surrounded by filaments that form a very young complex with characteristics very
similar to those of Orion \cite{wal91,barba}) suggests that the young cluster in SSC A could be 
initiated as a consequence of the energetic stellar activity of the older central cluster.

\vspace{5 cm}
Fig. 7: SED of the nucleus of NGC 7714 (left), fitted by composite model with contributions
 of bursts 5 Myr, 10 Myr and 200 Myr old; and NGC 604 (right) fitted by a 3 Myr old burst.
\vspace{1 cm}
%\begin{figure*}
%\psfig{figure=gonzalez_fig7.eps,width=210mm,angle=0}
%\caption{SED of the nucleus of NGC 7714 (left), fitted by composite model with contributions
% of bursts 5 Myr, 10 Myr and 200 Myr old; and NGC 604 (right) fitted by a 3 Myr old burst.} 
%\label{uvfit1} 
%\end{figure*}

\section{Summary}

RH\,{\sc ii}s and starbursts are objects in which the total energetics is 
dominated by star formation and associated phenomena. Their stellar content 
is responsible for their spectral morphology at ultraviolet and 
optical wavelengths. The ultraviolet is dominated by absorption lines formed
in the stellar wind of massive stars, and the optical by nebular emission
lines formed in the interstellar gas. However, at the Balmer jump, absorption
(mainly H and HeI photospheric lines) and nebular lines happen together. These 
characteristics emphasize the necessity of doing a multiwavelength analysis of 
these objects to derive their stellar content and evolutionary state in a 
consistent way. In particular, the analysis of the wind stellar and the nebular 
emission lines indicates
that the young stellar population in RH\,{\sc ii}s and starbursts lasts for
only a few Myrs. However, starburst galaxies are more complex star forming
systems than RH\,{\sc ii}s because their optical and near-infrared continuum 
are mainly produced by an intermediate age population. This suggests that the 
star formation in starbursts is taking place via several recurrent bursts 
lasting for a few hundred Myrs.

{\bf Acknowledgement}

I thank to my collaborators Marisa Garc\'\i a Vargas, Jeff Goldader, 
Claus Leitherer and Enrique P\'erez for their contributions to this work.

{\sc 

}

\end{document}